\documentclass[prd,aps,twocolumn,nofootinbib,superscriptaddress,eqsecnum,floatfix,longbibliography]{revtex4-1}

\usepackage{amssymb,amsmath}
\usepackage{graphicx}

\newcommand{\eps}{\varepsilon}
\newcommand{\tht}{\vartheta}
\newcommand{\ph}{\varphi}
\newcommand{\kap}{\varkappa}
\newcommand{\Alpha}{{\mathrm{A}}}
\newcommand{\Kappa}{{\mathrm{K}}}

\newcommand{\mo}{m_{\mathrm{o}}}
\newcommand{\qo}{q_{\mathrm{o}}}

\DeclareMathOperator\arctanh{arctanh}

\newcommand{\tens}[1]{{\boldsymbol{#1}}}

\newcommand{\be}{\begin{equation}}
\newcommand{\ee}{\end{equation}}

\newcommand{\srm}[1]{{\scriptscriptstyle\mathrm{#1}}}
\newcommand{\sss}[1]{{\scriptscriptstyle #1}}

\begin{document}

\title{Self-force on a static particle near a black hole}

\author{Pavel Krtou\v{s}}
\email{Pavel.Krtous@utf.mff.cuni.cz}
\affiliation{Institute of Theoretical Physics,
Faculty of Mathematics and Physics, Charles University,
V~Hole\v{s}ovi\v{c}k\'ach 2, Prague, Czech Republic}

\author{Andrei Zelnikov}
\email{zelnikov@ualberta.ca}
\affiliation{Theoretical Physics Institute, Department of Physics\\
University of Alberta, Edmonton, Alberta, Canada T6G 2E1}

\date{October 9, 2019}

\begin{abstract}
We study the self-force acting on a static charged point-like particle near a Schwarzschild black hole. We obtain the point-like particle as a limit of a spacetime describing a big neutral black hole with a small charged massive object nearby. The massive object is modeled by a black hole or a naked singularity. In this fully interacting system the massive object is supported above the black hole by a strut. Such a strut has a non-zero tension which corresponds to the external force compensating the gravitational force and the electromagnetic self-force acting on the massive object. We discuss details of the limiting procedure leading to the point-like particle situation. As a result, we obtain the standard gravitational force in the static frame of the Schwarzschild spacetime and the electromagnetic self-force. The electromagnetic self-force differs slightly from the classical results in a domain near the horizon. The difference is due to taking into account the influence of the strut on the electromagnetic field. We also demonstrate that higher order corrections to the gravitational force, a sort of gravitational self-force, are not uniquely defined, and they depend on details of the limiting procedure.
\end{abstract}
\maketitle



\section{Introduction}

The self-force problem is a problem of computing the motion of charged particles in an external gravitational field by taking into account a self-interaction of a particle with its own field. This problem has a long history going back to the classical works \cite{Abraham:1902,Lorentz:1915,Fermi:1921,DeWittBrehme:1960}. Even in flat spacetime it is not quite trivial to take into account radiation-reaction effects and the fact that an electromagnetic mass of a charged particle is not localized at a point. In curved spacetime the task becomes much more involved and subtle. In the four-dimensional case many approaches were developed for how to deal with this complicated problem. Three decades ago the study of self-force and self-energy was mostly related to the electric charges \cite{MoretteDeWitt:1964,Hanni:1973fn,McGruder:1978,Vilenkin:1979,Gibbons:1978,Smith:1980tv,Zelnikov:1982in,Zelnikov:1983,Lohiya:1982,Linet:1976sq,Leaute:1976sn,Linet:2000cv}. The perturbative approach was developed by Zerilli \cite{Zerilli:1971wd} and proved to be very effective in computations of the gravitational and electromagnetic self-force \cite{Bini:2006dp,Bini:2007zzd,Bini:2008zzd}. Currently, development in this field is mainly
motivated by the study of various processes in the vicinity of black holes or during black-hole collisions.

After the discovery of gravitational waves from binary black-hole mergers, a study of the self-force of compact objects in the black-hole background gained new interest. It can provide a very effective tool to test general relativity in a strong gravity limit. There are excellent reviews of the topic \cite{Poisson:2011nh,Pound:2015tma,Barack:2018yvs} where one can find a description of the contemporary methods, results, and applications of the self-force approaches discussed in the literature.

In four dimensions computation of the self-force in a strong gravitational field is technically quite involved, but conceptually, it is now well understood. It was surprising \cite{Beach:2014aba,Frolov:2014gla} that the generalization of the widely accepted and well-tested methods of computation of the self-energy
\cite{Frolov:2015ita,Frolov:2014fya,Frolov:2014kia,Frolov:2013qia,Frolov:2012ip,
Frolov:2012xf,Frolov:2012zd,Frolov:2012jj,Frolov:2014gla,Isoyama:2012in,
Kuchar:2013bla,Casals:2012qq,Zimmerman:2012zu,Zimmerman:2014uja,Ottewill:2012aj,
Galtsov:2007zz,LinzFriedmanWiseman:2014,Wald:2009ue,Gralla:2009uf,Gralla:2009md,Gralla:2008fg}
to higher-dimensional spacetimes led to unexpected ambiguities. It was noted
\cite{Beach:2014aba,Frolov:2014gla,Frolov:2015ita} that in odd-dimensional
spacetimes, the standard calculation of the self-force leads to some logarithmic terms depending on an unknown scale parameter. This problem appears even in a flat (Rindler) geometry \cite{Frolov:2014gla}. A few approaches were proposed \cite{Beach:2014aba,Frolov:2014gla,Taylor:2015waa,Harte:2016fru} for how to fix  these unpleasant ambiguities in higher dimensions. But although physically reasonable, they do not necessarily provide an invariant description of the self-force.

In four dimensions there remain some subtle problems with the invariant description of the gravitational self-force. In a first-order perturbation of the metric everything is clear: Any compact object moves as a test particle in the certain effective metric satisfying the linearized Einstein equations. Taking into account the self-force effects requires computations up to a second order of perturbations of the  geometry, which are non-linear. Nevertheless, a similar statement is still valid \cite{Pound:2017psq}, but it requires a refinement of a test point-particle approximation.

This approximation, like in electromagnetism, considers an extended object in the limit when mass, charge, momentum, and size scale to zero in a proportional manner \cite{Gralla:2008fg}. The self-force is given by the quadratic in these parameters. This limiting procedure leads to a physically satisfactory description of self-force effects for a test particle \cite{Gralla:2009md,Gralla:2009uf,Wald:2009ue,Mackewicz:2019sgm,Zimmerman:2012zu,Barack:2002bt,Broccoli:2018koi,Damour:2009sm}, but invariance of the self-force effects up to second order in these small parameters requires a special analysis \cite{Geroch:2017hdb,Barack:2018yvs}. We will return to this point at the end of the Introduction.

One of the ways to test our intuition and computational methods is to apply them to the exactly solvable models. This approach was successfully applied to study a self-force acting on a static charge in Schwarzschild spacetime \cite{MoretteDeWitt:1964,McGruder:1978,Vilenkin:1979,Gibbons:1978,Smith:1980tv,Zelnikov:1982in}, Kerr--Newman spacetime~\cite{Lohiya:1982}, Schwarzschild--de~Sitter spacetime \cite{Kuchar:2013bla}, and cosmic string spacetime \cite{Krtous:2006fb}, or to probe the spacetime global structure \cite{Davidson:2018vqe}. Let us recall that, for a particle at the static orbit in Schwarzschild spacetime, the ``classical'' answer for the magnitude of the electromagnetic self-force in the static frame is
\be\label{classEMsfforce}
F_{\text{sf}}= \frac{\qo^2 M}{r^3}\,,
\ee
and the self-force is pointing radially, away from the black hole \cite{Vilenkin:1979,Smith:1980tv,Zelnikov:1982in}.

In this paper we study the effect of a self-force exerted on a static electric charge placed in the vicinity of an uncharged black hole. The idea is to use the exact double black-hole solutions of the Einstein-Maxwell equations, where all back-reaction effects are taken into account exactly. Then we take the point-particle limit, when the mass and the charge of one component proportionally scale to zero. In this limit we obtain the test charged particle in the background geometry of the Schwarzschild black hole. We analyze the force which keeps the system in equilibrium.

The limiting system contains the black hole, the test charge, the electromagnetic field due to the charge, and the agent, which balances the particle at its position: the test strut between the black hole and the particle. The perturbation of the geometry due to the last term is frequently overlooked in some approaches. It is true that the strut is a test object which does not affect the limiting Schwarzschild geometry. However, the self-force of a test charge is the second-order perturbation effect. The test strut or string influences the geometry  and modifies the surrounding electromagnetic field, which in turn affects the self-force.\footnote{%
A similar approach has been adopted in the work of LaHaye and Poisson \cite{LaHaye:2020lsb}, which appeared during the publication process of this paper.}

In our model of the fully interacting system, two black holes are kept in equilibrium by a strut described by a conical defect between them. The geometry and, consequently, the electromagnetic field are affected by the presence of this conical defect. We compute the limiting self-force on the particle by taking into account this effect.  As a result, we obtain
\begin{equation}\label{ourEMsfforce}
    F_{\text{sf}} = \frac{\qo^2 M}{r^3} \frac{1-\frac{M}{r}}{1-\frac{2M}{r}}\,.
\end{equation}
This expression differs from the classical result \eqref{classEMsfforce}. The reason for this difference is that it also takes into account the influence of the strut which keeps the particle at its position. Both expressions for the self-force agree sufficiently far from the horizon, but they differ near the horizon, when the tension and the linear energy density of the strut are big.

Finally, let us comment on the well-definiteness of our limiting procedure. Limits of the spacetimes are rather involved and depend on many ingredients \cite{Geroch:1969ca,Pound:2015fma}. The spacetime description itself must deal with the standard diffeomorphism freedom, and the limiting procedure adds additional ambiguity. For example, in our system of a test particle near the black hole, even the answer to the simple question ``What position of the black hole corresponds to a position of the point particle?'' depends on the limiting procedure.

In Ref.~\cite{Pound:2015fma} three types of ``gauge freedom'' have been described: (i) identification of points in the perturbed spacetime with points in the background spacetime; (ii) the freedom to perform a coordinate transformation (or diffeomorphism) on the background, inducing, via an identification map, a coordinate transformation in the full spacetime; (iii) the choice of the spacetime family used in the limiting procedure. This freedom corresponds to a choice of ``what to hold fixed'' while taking the limit.

The first two freedoms are related to the standard diffeomorphism gauge freedom of general relativity, and we do not discuss them in much detail. In our paper, we fix the identification of points during the limit using the Weyl coordinates. Other choices could lead to diffeomorphically equivalent results---if the identification of points does not differ significantly; or one obtains qualitatively different results---if the identification of points during the limiting procedure does, for example, infinitely zoom in on some particular area or stretch some particular direction.\footnote{%
Typical examples of such a dependence on the point identification are near-horizon limits of black-hole geometries.}

In this paper, we are not interested in these phenomena. We are primarily concerned with the third type of freedom above. As discussed in \cite{Pound:2015fma}, this freedom, in general, cannot be removed or compensated by small coordinate transformations. This freedom reflects the choice of a particular limiting procedure. The choice is fixed by coordinate independent physical observables that are used to identify points of a set of different spacetimes. Different gauge-invariant limiting procedures may lead to physically different predictions for the gravitational self-force of a test particle, while the electromagnetic self-force is independent of these choices. The position of the test particle is defined by the physical observables, like proper distance, thermodynamic length, red shift factor, and other invariant observables. In the literature, though, it is common to fix the distance from the black hole as a coordinate distance, e.g., Schwarzschild or Weyl radial coordinate. Mathematically, it is a very convenient gauge choice, especially for perturbative computations, but implicitly, it assumes a very particular gauge-noninvariant limiting procedure.

This ambiguity may affect the expressions for the self-energy, the self-force, and the effective equations of motion of the particle. It is important to find out which effects and quantities are robust and which ones depend on the limiting procedure. This freedom can be fixed by choosing operationally defined quantities that are held fixed while taking the limit. There may be several physically motivated operational choices for these quantities, and it is not guaranteed that they will provide the same prediction for the self-force. Of course, if one makes the choice of the limiting procedure, i.e., the choice of experimentally measurable observables which are kept fixed, then the prediction for the self-force becomes unambiguous.

We demonstrate the robustness of the electromagnetic self-force, i.e., its independence on the limiting procedure. As for the gravitational self-force, we find that it does depend on the choice of variables that are kept fixed in the limiting procedure. For example, keeping the total mass measured at spatial infinity constant gives a different gravitational self-force than when keeping the surface gravity of the black hole fixed. Depending on the physical motivations and further applications of the result, one can thus have equally well justified but different expressions for the gravitational self-force.

The plan of our work is as follows: First, in Sec.~\ref{sc:2BHsol}, we describe the system of two black holes and review its geometrical and thermodynamical characteristics. In Sec.~\ref{cs:test-charge}, we briefly recall the concept of the test charge and previously known results for the electromagnetic self-force. In Sec.~\ref{sc:SFbylimit}, we describe the limiting procedure to a point test particle, derive the gravitational and electromagnetic self-forces, and discuss their properties. In Sec.~\ref{sc:Summary}, we summarize and discuss our results.


\section{Double black-hole solution}
\label{sc:2BHsol}


\subsection{Geometry}
\label{ssc:geometry}

An asymptotically flat static solution of Einstein-Maxwell equations describing two nonextreme charged black holes in equilibrium was obtained in \cite{Alekseev:2007gt,Alekseev:2007re,Alekseev:2011nj,Manko:2007hi,Manko:2008gb}.  Its metric and the electromagnetic vector potential can be written in cylindrical Weyl coordinates as
\be\label{metric}
ds^2=-f\,dt^2+f^{-1}\left[h^2(d\rho^2+dz^2)+\rho^2\,d\varphi^2\right],
\ee
\be
A_t=-\Phi\,,\quad A_\rho=A_z=A_\varphi=0.
\ee
Here $f,h$ and $\Phi$ are the functions of the coordinates $\rho$ and $z$ only.

\begin{figure}[t]
\centering
\includegraphics[width=8.5cm]{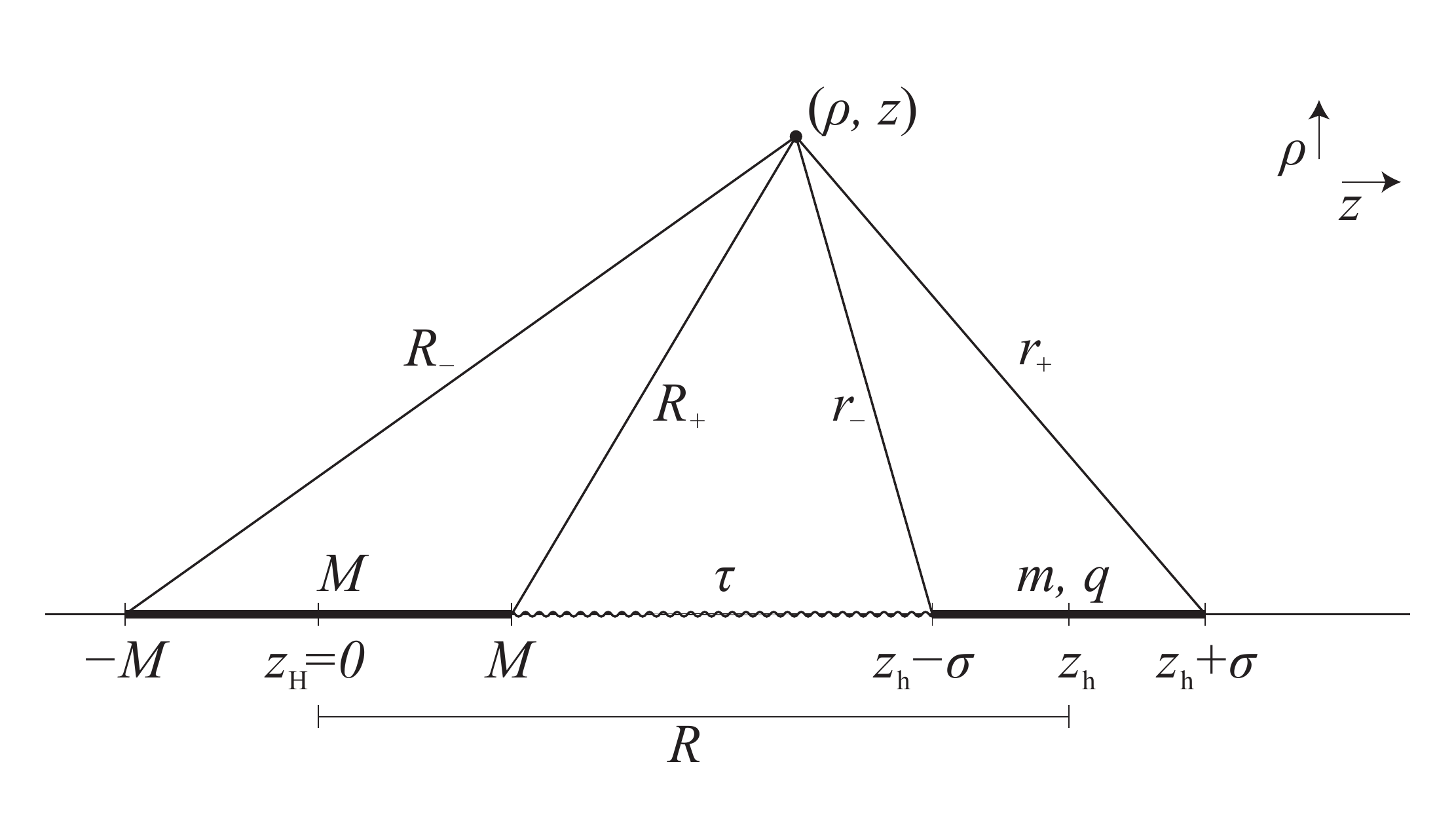}
  \caption{The spacetime of two back holes in Weyl coordinates ${\rho}$,~${z}$, (${t=\text{const}}$, ${\varphi=\text{const}}$). The horizons of both holes are squeezed into coordinate-singular rods placed on the symmetry axis $\rho=0$. The centers of these rods are at $z_h$ and $z_H$, the half-lengths of the rods are ${\sigma}$ and ${\Sigma}$. The quantities ${r_{\sss\pm}}$ and ${R_{\sss\pm}}$ are evaluated as coordinate distances from the ends of the rods. They play a role in the expressions for the metric functions. The figure is adjusted to the case investigated in this paper, when one of the black holes is uncharged, ${Q=0}$ (which implies ${\Sigma=M}$), and it is placed at the origin, ${z_H=0}$.
\label{fig:BHBHWeyl}}
\end{figure}

The Weyl coordinates describe the exterior of the black holes. Horizons of both black holes degenerate into two rods localized on the axis ${\rho=0}$, see Fig.\,\ref{fig:BHBHWeyl}. The centers of these rods are localized at ${z_H}$ and ${z_h}$, respectively and their coordinate distance is given by the separation parameter ${R}$
\be
R=|z_h-z_H|.
\ee
Without loss of generality, we can set the origin of the ${z}$ coordinate such that ${z_H=0}$.
The half-lengths $\Sigma$ and $\sigma$ of the rods are given by
\be\label{sigmas}
\Sigma^2=M^2-Q^2+2\mu Q\,,\quad
\sigma^2=m^2-q^2-2\mu q\,.
\ee
where ${M,\,m}$ and ${Q,\,q}$ are masses and charges of the black holes, respectively.
Here and below we use several constants:
\be\begin{gathered}\label{constsdef}
  \mu=\frac{mQ-Mq}{R+M+m}\;,\\
  \nu=R^2-\Sigma^2-\sigma^2+2\mu^2\,,\\
  \kap=M m -(Q-\mu)(q+\mu)\,,\\
  K_*=4\Sigma\sigma\bigl(R^2-(M-m)^2+(Q-q-2\mu)^2\bigr)\,.
\end{gathered}\ee

The metric functions ${f}$ and ${h}$ are
\be\label{fh}
  f=\frac{\mathcal{A}^2-\mathcal{B}^2+\mathcal{C}^2}{(\mathcal{A}+\mathcal{B})^2}\,,\quad
  h^2=\frac{\mathcal{A}^2-\mathcal{B}^2+\mathcal{C}^2}
      {K_*^2 R_{\sss+} R_{\sss-} r_{\sss+} r_{\sss-}}\,,
\ee
and the potential for the Maxwell field is
\be\label{Phi}
\Phi=\frac{\mathcal{C}}{\mathcal{A}+\mathcal{B} }\;.
\ee
Here, $\mathcal{A}$, $\mathcal{B}$, $\mathcal{C}$ are complicated auxiliary functions \cite{Alekseev:2007re,Alekseev:2011nj,Manko:2007hi,Manko:2008gb} which are listed in Appendix~\ref{apx:mtrcfcs}. They are written in terms of coordinate distances from the endpoints of the rods, cf.~Fig.\,\ref{fig:BHBHWeyl}:
\be\label{coordists}\begin{aligned}
R_{\sss\pm}&=\sqrt{\rho^2+(z-z_H\mp\Sigma)^2}\,,\\
r_{\sss\pm}&=\sqrt{\rho^2+(z-z_h\mp\sigma)^2}\,.
\end{aligned}\ee

The solution describes two black holes when the quantities ${\sigma}$ and ${\Sigma}$ are real and satisfy the condition
\begin{equation}\label{BHcond}
    R>\Sigma+\sigma\,.
\end{equation}
The equality in this condition would represent the limit of touching black holes. In the uncharged case, ${Q,q=0}$, it leads to the spacetime with one black hole of mass~${M+m}$.

\begin{figure}[t]
\centering
\includegraphics[width=8.5cm]{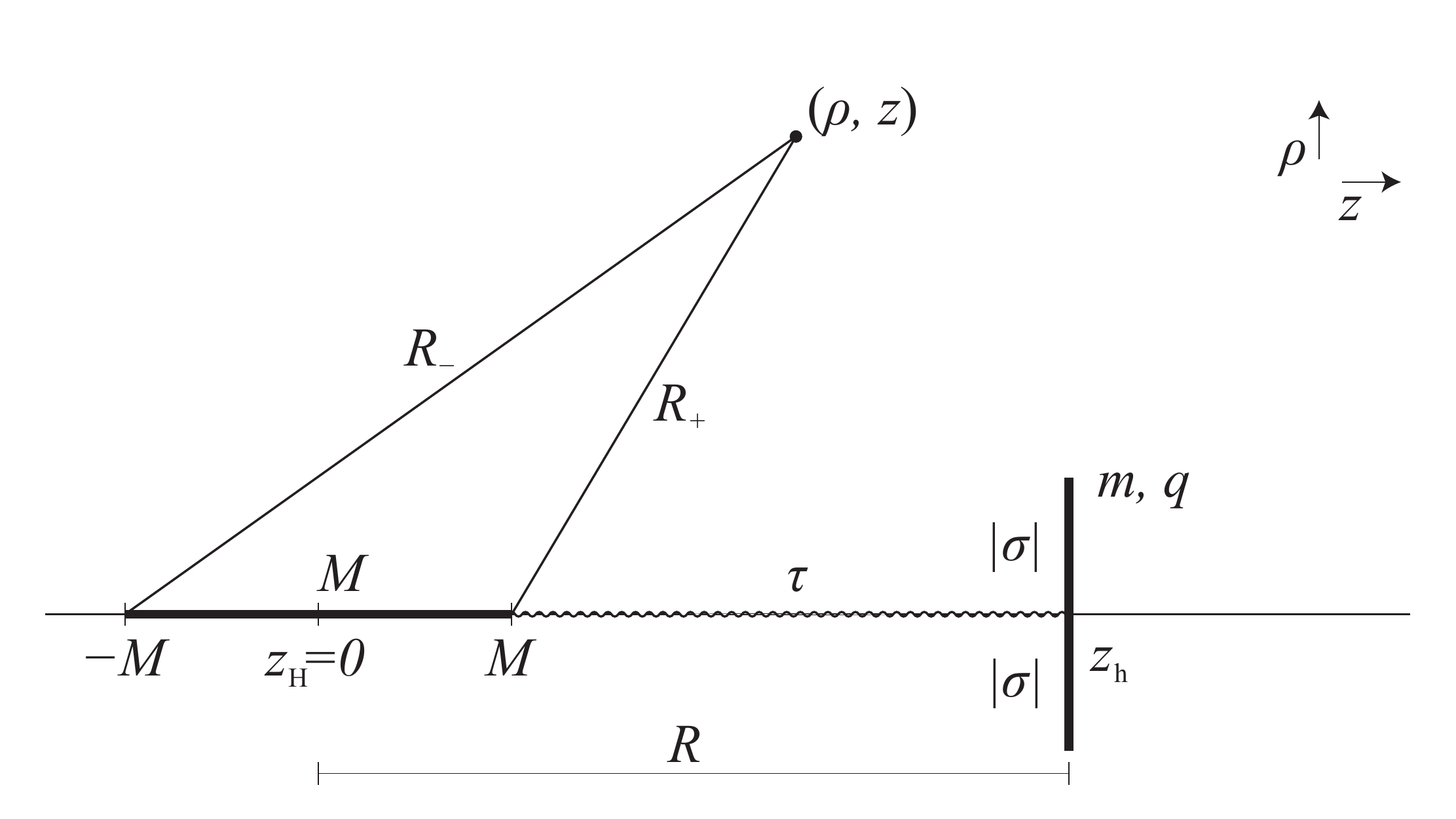}
  \caption{Spacetime of an uncharged black hole with a charged naked singularity nearby, shown in Weyl coordinates ${\rho}$,~${z}$, (${t=\text{const}}$, ${\varphi=\text{const}}$). The horizon of the black hole is represented by the rod placed on the symmetry axis $\rho=0$ at ${z=z_H\equiv0}$. At the coordinate ${z=z_h\equiv R}$, positioned symmetrically around the axis, there is a disk-like coordinate singularity of the radius ${|\sigma|}$. It plays the role of a wormhole: the spacetime can be extended through the disk by gluing another patch of Weyl coordinates with analytically extended metric functions. At the coordinate ${m}$ behind the wormhole along the symmetry axis there is a naked curvature singularity. The quantities ${R_{\sss\pm}}$ are the coordinate distances from the ends of the rod. Similar quantities ${r_{\sss\pm}}$ associated with the naked singularity are complex, however, they still lead to the real metric functions.
\label{fig:BHSWeyl}}
\end{figure}

Negative ${\Sigma^2}$ and/or ${\sigma^2}$ correspond to the presence of naked singularities instead of black holes \cite{Alekseev:2007gt}. Although in this case ${\Sigma}$ and/or ${\sigma}$ are imaginary, and, as a consequence, ${R_{\sss\pm}}$ and/or ${r_{\sss\pm}}$ are complex (but complex conjugate to each other), the metric functions turn out to be real as needed for a physical interpretation.

In the black-hole case, the Weyl potential ${\frac12\log f}$ is, in fact, the Newtonian potential corresponding to the two horizon rods as sources with density ${\frac12}$ in an auxiliary Euclidian space covered by the cylindrical coordinates ${\rho,\,\ph,\,z}$. In the naked singularity case the rods are replaced by disks positioned at ${z_H}$ and/or ${z_h}$ of the radii ${|\Sigma|}$ and/or ${|\sigma|}$, respectively, cf.~Fig.~\ref{fig:BHSWeyl}.

Both rods and disks are just coordinate singularities of the Weyl coordinates. The rods correspond to horizons, and the spacetime can be extended through them into the interior of black holes. The disks are just branch surfaces of the Weyl coordinates, and they can be viewed as wormholes into other parts of the spacetime. At each disk one needs to glue the Weyl coordinates with another patch of coordinates on the other side of the disk. Such a patch describes a prolongation of the spacetime. When following the symmetry axis through the disk, at the coordinate distance ${M}$ (or ${m}$ for the other disk), one encounters a naked curvature singularity \cite{Alekseev:2007gt}.


\subsection{Double black-hole solution. Physical quantities}\label{ssc:physquant}

The described solution has been thoroughly analyzed in \cite{Alekseev:2007gt,Alekseev:2007re,Alekseev:2011nj,Cabrera-Munguia:2018omi,Manko:2007hi,Manko:2008gb}. The most physically interesting quantities have been calculated, and we just list them here. The formulas assume the black-hole case. But typically, the quantity defined for one of the black holes remains well defined when the other black hole is changed to a naked singularity. In this case one has to remember that the corresponding ${\Sigma^2}$ or ${\sigma^2}$ is negative.

The total mass of the system is
\begin{equation}\label{energy}
    \mathcal{M} = M+m\;.
\end{equation}
The areas of the horizons of both black holes are
\be\label{areas}
\begin{aligned}
    A&=4\pi\frac{\bigl((R+M+m)(M+\Sigma)-Q(Q+q)\bigr)^2}{(R+\Sigma)^2-\sigma^2}\,,
    \\
    a&=4\pi\frac{\bigl((R+M+m)(m+\sigma)-q(Q+q)\bigr)^2}{(R+\sigma)^2-\Sigma^2}\,,
\end{aligned}
\ee
the surface gravities are
\be\label{surfgrs}
\begin{aligned}
    \Kappa&=\frac{\Sigma\,\bigl((R+\Sigma)^2-\sigma^2\bigr)}
       {\bigl((R+M+m)(M+\Sigma)-Q(Q+q)\bigr)^2}\,,
    \\
    \kappa&=\frac{\sigma\,\bigl((R+\sigma)^2-\Sigma^2\bigr)}
       {\bigl((R+M+m)(m+\sigma)-q(Q+q)\bigr)^2}\,,
\end{aligned}
\ee
and the electric potentials on the horizons are
\be\label{potentials}
    \Phi=\frac{Q-2\mu}{M+\Sigma}\,,
    \quad
    \phi=\frac{q+2\mu}{m+\sigma}\,.
\ee

The total charges of each black hole are ${Q}$ and ${q}$, respectively. It is not a simple task to identify a mass of each black hole separately since one cannot avoid the nonlinear nature of the mutual interaction. But it is argued in \cite{Alekseev:2007re} that the parameters ${M}$ and ${m}$ directly describe the individual masses of the black holes. One can also observe a remarkable property that both of these parameters satisfy the Smarr relations in the form
\begin{equation}\label{sepSmarr}
    M = 2 T S + \Phi Q\,,
\quad
    m = 2 t s + \phi q\,,
\end{equation}
where entropies ${S}$, ${s}$ and temperatures ${T}$, ${t}$ are defined in the standard way,
\begin{gather}
    \label{entropydef}
    S = \frac{A}{4}\,,
    \quad
    s = \frac{a}{4}\,,\\
    \label{tempdef}
    T = \frac{\Kappa}{2\pi}\,,
    \quad
    t = \frac{\kappa}{2\pi}\,.
\end{gather}

Both black holes (or naked singularities) interact, besides through the gravitational and electromagnetic interaction, also through a strut localized on the axis between them. It can be shown that the axis between black holes is not smooth but contains a conical singularity. Such a singularity represents a thin physical source with an internal energy and a tension. These can be related to the conical defect on the axis \cite{Israel:1976vc,Alekseev:2007re,Manko:2007hi}. When the angle ${\Delta\phi}$ around the axis is smaller than the full angle ${\Delta\phi = 2\pi-\delta}$, with $\delta>0$,  the object on the axis is called the cosmic string. If the angle around the axis is bigger than $2\pi$, then $\delta<0$, and the object represents the strut \cite{Israel:1976vc}. The strut has a negative energy density ${\eps}$ and a positive linear pressure $\tau$, which is also called the tension of the strut. These are related to the angular excess ${-\delta>0}$ as ${\tau=-\eps=-\frac{\delta}{8\pi}}>0$. Intuitively, because of the equality between linear energy density and tension, the effective gravitational masses of the string or the strut vanish. As a consequence, the influence on the surrounding spacetime is special: it effectively causes only the conical defect on the axis.

The system discussed contains a strut between the black holes with the tension \cite{Manko:2007hi}
\begin{equation}\label{tension}
    \tau =\frac{\kap}{\nu-2\kap}=\frac{Mm-(Q-\mu)(q+\mu)}{R^2-(M+m)^2+(Q+q)^2} \,.
\end{equation}
One can also associate with the strut a conjugate thermodynamical observable called the thermodynamic length ${\ell}$, see \cite{Krtous:2019fpo}. It has the meaning of the strut worldsheet area per unit of the Killing time,
\begin{equation}\label{elldef}
\ell = \frac{1}{\Delta t}\int_{\text{strut}}d\mathfrak{A}
     =\frac{1}{\Delta t}\int_t^{t+\Delta t} \!\!\!\int_z h\big|_{\rho=0}\, dz\, dt\;.
\end{equation}
The metric function ${h}$ is constant on the axis and between the black holes (or naked singularities). It has the value
\be\label{honaxis}
h\big|_{\rho=0}=h_{\mathrm{o}}\equiv\frac{\nu-2\kap}{\nu+2\kap}\,.
\ee
For the case of two black holes one integrates over the part of the axis between the horizons and the thermodynamic length is thus
\be\label{ellHh}
\begin{split}
\ell_{\srm{Hh}} &=(R-\Sigma-\sigma)\frac{\nu-2\kap}{\nu+2\kap}\\
    &=(R-\Sigma-\sigma)\frac{R^2-(M+m)^2+(Q+q)^2}{R^2-(M-m)^2+(Q-q-2\mu)^2}\,.
\end{split}\raisetag{9ex}
\ee
For the case of a naked singularity of mass ${m}$ and charge ${q}$ near the black hole of mass ${M}$ and charge ${Q}$ one integrates between the horizon and the singularity, yielding
\be\label{ellHs}
\begin{split}
\ell_{\srm{Hs}} &=(R-\Sigma+m)\frac{\nu-2\kap}{\nu+2\kap}\\
    &=(R-\Sigma+m)\frac{R^2-(M+m)^2+(Q+q)^2}{R^2-(M-m)^2+(Q-q-2\mu)^2}\,.
\end{split}\raisetag{9ex}
\ee

The proper length of the strut is typically more complicated, and it is evaluated in Appendix~\ref{apx:PropLength}.

Either of the black holes or both can be made extremal independently. This happens when $\Sigma^2=0$ and/or $\sigma^2=0$. The corresponding conditions for the charges are
\be\label{qextr}
\begin{aligned}
Q&=\frac{M}{R{+}M{-}m}\Bigl(\pm\sqrt{(R{+}M)^2{+}q^2{-}m^2}-q\Bigr)\,,
\\
q&=\frac{m}{R{+}m{-}M}\Bigl(\pm\sqrt{(R{+}m)^2{+}Q^2{-}M^2}-Q\Bigr)\,.
\end{aligned}
\ee
Of course, the extremal case corresponds to the boundary at which the black-hole spacetime changes into the naked singularity spacetime.


\subsection{Neutral black hole}
\label{ssc:unchargedBH}

In the following sections we study a small black hole or naked singularity near a big neutral black hole of mass ${M}$. In this case ${Q=0}$, ${\Sigma=M}$, and the thermodynamic quantities reduce to
\be\label{TDQ0}
\begin{aligned}
    S&=4\pi M^2 \frac{(R+M+m)^2}{(R+M)^2-\sigma^2}\,,
    \\
    s&=\pi\frac{\bigl((R+M+m)(m+\sigma)-q^2\bigr)^2}{(R+\sigma)^2-M^2}\,,
    \\
    T&=\frac{1}{8\pi M}\frac{(R+M)^2-\sigma^2}{(R+M+m)^2}\,,
    \\
    t&=\frac{1}{2\pi}\frac{\sigma\,\bigl((R+\sigma)^2-M^2\bigr)}
       {\bigl((R+M+m)(m+\sigma)-q^2\bigr)^2}\,.
\end{aligned}
\ee
The potentials on the horizons are
\be\label{potentialsQ0}
    \Phi=\frac{q}{R+M+m}\,,
    \quad
    \phi=\frac{q}{m+\sigma}\frac{R-M+m}{R+M+m}\,.
\ee
The tension of the strut has the form
\begin{equation}\label{tensionQ0}
    \tau =\frac{\kap}{\nu-2\kap}=\frac{Mm-(Q-\mu)(q+\mu)}{R^2-(M+m)^2+(Q+q)^2} \,.
\end{equation}
The thermodynamic length in the case of a small black hole reduces to
\be\label{ellHhQ0}
\ell_{\srm{Hh}}
    =\frac{(R{-}M{-}\sigma)\bigl(R^2-(M{+}m)^2+q^2\bigr)}
      {R^2-(M{-}m)^2+q^2\frac{(R{-}M{+}m)^2}{(R{+}M{+}m)^2}}\,,
\ee
and in the case of a small naked singularity to
\be\label{ellHsQ0}
\ell_{\srm{Hs}}
    =\frac{(R{-}M{+}m)\bigl(R^2-(M{+}m)^2+q^2\bigr)}{R^2-(M{-}m)^2+q^2\frac{(R{-}M{+}m)^2}{(R{+}M{+}m)^2}}\,.
\ee

The extremality condition \eqref{qextr} for a small black hole yields
\be
q^2=m^2\frac{R+M+m}{R-M+m}\,.
\ee
For the square of charge ${q^2}$ smaller than this critical value, the spacetime describes two black holes; if ${q^2}$ is larger, it represents a charged naked singularity above the uncharged black hole.


\subsection{Schwarzschild geometry}
\label{ssc:Schw}

For ${Q=0}$, ${m=0}$, and ${q=0}$, the geometry reduces to the Schwarzschild solution of mass ${M}$. In the Weyl coordinates it has the form given by the metric functions
\begin{equation}\label{Schw}
    f = \frac{R_{\sss+}{+}R_{\sss-}{-}2M}{R_{\sss+}{+}R_{\sss-}{+}2M}\,,\quad
    h^2 = \frac{(R_{\sss+}{+}R_{\sss-})^2-4M^2}{4R_{\sss+}R_{\sss-}}\;.
\end{equation}
The transformation from the Weyl coordinates ${t,\,\rho,\,z,\,\ph}$ to the Schwarzschild spherical coordinates ${t,\,r,\,\tht,\,\ph}$ is \cite{Griffiths:2009dfa,Griffiths:2006tk}
\begin{equation}\label{WeylSchwTr}
   \rho = \sqrt{r(r-2M)}\,\sin\tht\,,\quad
   z = (r-M) \cos\tht\;.
\end{equation}
In particular, along the semiaxis ${\tht=0}$, i.e., ${\rho=0}$, ${z>0}$, we have
\begin{equation}\label{zronaxis}
    r=z+M\,.
\end{equation}


\section{Self-force of a test charge}\label{cs:test-charge}

A test charged particle in a gravitational field, i.e., in a curved spacetime, creates an electromagnetic field in the spacetime. For an extended object, such a field interacts with the object itself. Therefore, one can expect that in the limit of a point particle, such an interaction survives in the form of a self-force. The self-force acts on the generically moving point particle already in Minkowski spacetime \cite{Abraham:1902,Lorentz:1915,Fermi:1921}. This interaction can be understood as a reaction on the field radiated by the particle. The self-force also can be evaluated in the curved spacetime \cite{DeWittBrehme:1960,Quinn:1996am,Mino:1996nk}, where there are additional contributions due to scattering of the electromagnetic field on the curvature.

For a static charged particle in the Schwarzschild or Reissner-Nordstrom spacetimes, the electromagnetic self-force has been evaluated by various methods, see, e.g., Refs.~\cite{MoretteDeWitt:1964,McGruder:1978,Vilenkin:1979,Gibbons:1978,Smith:1980tv,Zelnikov:1982in}. We phrase the results in terms of the external force which is needed to support the particle at the static orbit. The total force ${\tens{F}_{\text{ext}}= F_{\text{ext}} \tens{e}_r}$ needed to support the test particle of a rest mass ${\mo}$ and of a charge ${\qo}$ floating at the Schwarzschild radius ${r}$ near the black hole of mass ${M}$ is
\be\label{classforce}
F_{\text{ext}}=\frac{\mo M}{r^2 }\Bigl(1-\frac{2M}{r}\Bigr)^{\!-\frac12}-\frac{\qo^2 M}{r^3}\,,
\ee
where ${\tens{e}_r}$ is the normalized radial vector in the static and locally comoving frame.
The first term balances a classical gravitational force in the static frame at radius~${r}$. The second term is equal to the additional self-force due to the self-interaction of the charged particle with its own electromagnetic field. The characteristic of the self-force is that it is proportional to a square of the charge and to the mass of the black hole, and it always points away from the black hole.

In principle, there exists a self-floating solution when the self-force exactly balances the gravitational force. However, it occurs only for unphysical values of the involved quantities, namely, for the black hole with a gravitational radius smaller than the classical ``radius'' of the point particle ${\frac{q^2}{\mo}}$ and at a distance comparable with this radius, see discussion, e.g., in \cite{Smith:1980tv}.

Similarly to the electromagnetic self-force, one could expect that the point particle acts on itself also through the gravitational self-force. To estimate such an interaction, however, is a much more difficult task since it involves an evaluation of the backreaction of the singular source on the spacetime geometry, which, due to the nonlinear nature of the Einstein equation, is not an easily defined problem. However, there is a wide variety of approaches to this problem in the recent literature (see, e.g., Refs.~\cite{Poisson:2011nh,Pound:2015tma,Barack:2018yvs} and references therein).

In various approaches, a common feature of the gravitational self-force is that it is not as unambiguous as the electromagnetic self-force. It usually depends on details of how the self-force is evaluated and how the approximation of the point-like particle is obtained.


\section{Self-force from a limit of a fully backreacting system}
\label{sc:SFbylimit}


\subsection{Limiting procedure}
\label{ssc:limit}

In our approach of evaluating the self-force acting on the static point-like particle in the Schwarzschild geometry, we start with an exact solution of the Einstein-Maxwell equations representing a big uncharged black hole of mass ${M}$ and a small massive object of mass ${m}$ and charge ${q}$, which can be either a small black hole or a naked singularity, depending on the values of ${m}$ and ${q}$. Such a solution has been described in Sec.~\ref{sc:2BHsol}.

This solution contains all of the information about the gravitational and electromagnetic interaction between a big black hole and a small massive object, including all kinds of gravitational and electromagnetic ``self-interaction''. It also describes the agent which keeps both objects in a static equilibrium, namely, the strut localized on the axis between the objects. This strut has a linear energy ${\eps}$ and a linear pressure (tension) ${\tau}$ along the axis. This pressure exactly corresponds to the external force which is needed to keep the massive object at a constant distance above the black hole.

Next we perform a limit in which the mass and charge of the massive object become small and the massive object changes into a point-like test particle. The strut also becomes a test source, which no longer influences the resulting background geometry. However, it still has a tension which corresponds to the external force needed to support the test particle at the static orbit and it balances both the gravitational and electromagnetic interactions.

\begin{figure}[t]
\centering
\includegraphics[width=8.5cm]{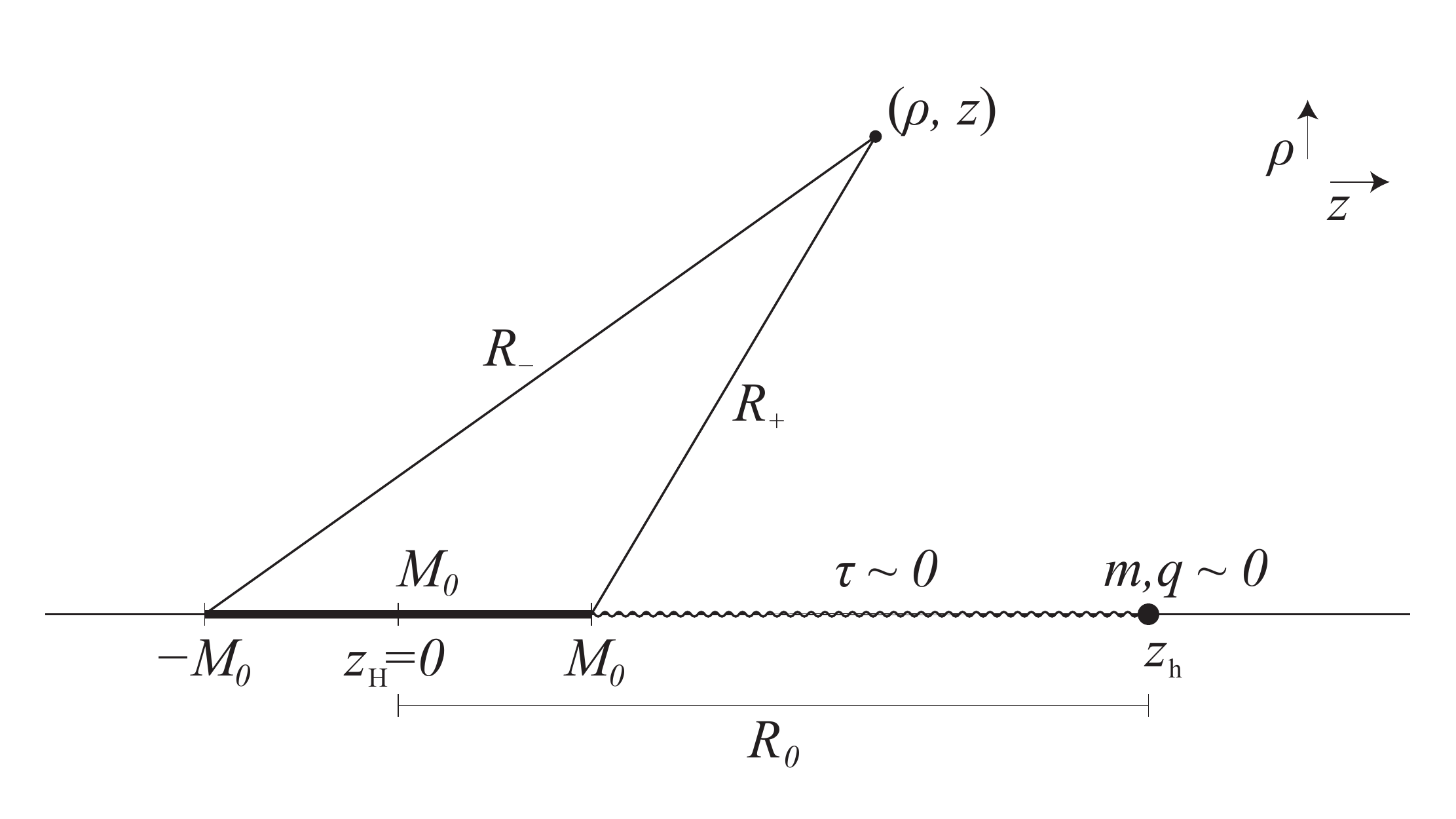}
  \caption{\label{fig:BHPWeyl}
  Limiting Schwarzschild spacetime in Weyl coordinates. The horizon of the neutral black hole of mass ${M_0}$ is represented by the rod of the half-length ${M_0}$. The point-like particle of a test mass ${m}$ and charge ${q}$ is localized at ${z=R_0}$, ${\rho=0}$. It is a remnant of a small black hole (cf.~Fig.~\ref{fig:BHBHWeyl}) or a naked singularity (cf.~Fig.~\ref{fig:BHSWeyl}) in the limit ${m,q\to0}$, i.e., when the rod or the disk representing the black hole or naked singularity, respectively, shrinks to a point.
}
\end{figure}

As already mentioned, we perform the limit in the class of double black-hole spacetimes characterized by parameters ${M}$, ${R}$, ${m}$, and ${q}$. We know that for ${m=0}$, ${q=0}$, the geometry reduces to the Schwarzschild geometry, Fig.~\ref{fig:BHPWeyl}. This means that we need to approach the values  ${[M_0,\,R_0,\,0,\,0]}$ with a curve ${[M(\epsilon),\,R(\epsilon),\,m(\epsilon),\,q(\epsilon)]}$ in the parametric space, where ${M_0}$ and ${R_0}$ are just limiting values of the mass of the big black hole and of the separation parameter.

However, to identify the position of the test particle in the final Schwarzschild geometry of mass ${M_0}$, it is necessary to identify points of the manifolds during the limiting process. It is well known \cite{Geroch:1969ca} that different identifications can lead to different limiting spacetimes. Indeed, the suitably chosen identification of points can incorporate zooming of some parts of the spacetime and squeezing of others.

In our procedure we identify points by fixing the Weyl coordinates during the limiting process. The points for different values of the spacetime parameters are identified if they have the same Weyl coordinates. Of course, this defines the identification only in the static domain outside the black holes, but it is the domain which we are interested in. We also assume that the big uncharged black hole is localized at ${z_H=0}$ during the limiting procedure.\footnote{This just fixes one degree of the diffeomorphism freedom.}

With such an identification, as a result of the limit, the small black hole or naked singularity reduces to a point-like object localized on the axis at ${z=z_h\equiv R_0}$ in the Weyl coordinates. In the Schwarzschild coordinates this corresponds to
\begin{equation}\label{partposr}
    r = R_0 + M_0\,,
\end{equation}
cf.\ relation~\eqref{zronaxis}.


\subsection{Limit ${m,\,q\to 0}$ with ${m\sim q}$}
\label{ssc:limitmsimq}

Now we have to stipulate in more detail, how we approach the limiting spacetime. For that we specify an expansion of the parametric curve ${[M(\epsilon),\,R(\epsilon),\,m(\epsilon),\,q(\epsilon)]}$ near its limiting value ${\epsilon=0}$,
\begin{equation}\label{mqExpansion}
\begin{aligned}
    m(\epsilon) &= \hat{m}\,\epsilon\,,\\
    q(\epsilon) &= q_1\,\epsilon + q_2\, \epsilon^2 + \dots\,,\\
    M(\epsilon) &= M_0 + M_1\, \epsilon + \dots\,,\\
    R(\epsilon) &= R_0 + R_1\, \epsilon + \dots\,.\\
\end{aligned}
\end{equation}
It is essentially the limit in small mass ${m}$, and we require that the charge scales to zero as well. We assume that the mass and charge of the massive object approach zero in the same order.  Therefore, the massive object can represent both a black hole and a naked singularity during the limiting procedure.

By setting coefficients ${q_1,\,q_2\,,\dots}$ to zero, we have the case in which we shrink a small neutral black hole. The case of a naked singularity with charge ${q}$ much larger than mass ${m}$ will be explored in the next subsection.

We have kept the higher order coefficients in expansions \eqref{mqExpansion} to have control over details of the limiting procedure. This is because we still need to specify, based on physical grounds, how we should perform the limit. It is natural to require that we perform the limit by keeping the big hole and its separation from the massive object ``unchanged.'' However, the spacetime changes during the limit, so we do not expect that the big black hole remains completely unchanged. We can choose a particular characteristic which remains the same in the limiting procedure.

A natural candidate is the mass ${M}$ of the black hole. But one could also consider the entropy~${S}$ (the area) of the black hole, or temperature ${T}$ (the horizon surface gravity), or maybe the total mass ${\mathcal{M}}$ of the system.

For the separation of the massive object from the black hole, the situation is even more ambiguous. We can keep the separation parameter ${R}$ constant, but this does not have a direct physical meaning---it is a coordinate distance between fictitious centers of the black holes. A more plausible choice for two black holes could be to keep the coordinate distance between the horizons, ${R-\Sigma-\sigma}$, constant. For a naked singularity near the black hole one could consider the coordinate separation up to the singularity: ${R+m}$ from the black hole ``center'' or ${R-\Sigma+m}$ from the horizon. Moreover, instead of the coordinate separation, it would be more natural to use the thermodynamic length ${\ell}$ or the proper distance ${L}$. All of these choices define different limiting curves in the parametric space. Therefore, we have to investigate whether this choice influences the resulting force acting on the test particle.

For that we need to expand the tension \eqref{tensionQ0} along the limiting curve. See Appendix~\ref{apx:expansions} for expansions of some intermediate quantities. Here, we just list a leading term of ${\sigma}$ for further reference,
\begin{equation}\label{sigma1}
    \sigma = \sigma_1\,\epsilon+\dots\,,\quad
    \sigma_1 = \sqrt{\hat{m}^2-q_1^2\, \frac{R_0{-}M_0}{R_0{+}M_0}}\,,
\end{equation}
The expansion of the tension is
\begin{equation}\label{tensionexp}
\begin{split}
    \tau &= \frac{\hat{m}M}{r^2}\Bigl(1-\frac{2M}{r}\Bigr)^{\!-1}\epsilon\\
     &\quad+\frac{2\hat{m}^2 M^2
     + \hat{m}\Bigl(M_1 r^2 - 2M (M_1{+}R_1)(r{-}M)\Bigr)}{r^4\bigl(1-\frac{2M}{r}\bigr)^2}\,
     \epsilon^2\\
     &\quad-\frac{q_1^2 M}{r^3} \frac{1-\frac{M}{r}}{1-\frac{2M}{r}}\,\epsilon^2
     + \dots \,.
\end{split}\raisetag{6ex}
\end{equation}
Here, for readability reasons, we have changed the final mass ${M_0}$ and the separation parameter ${R_0}$ to ${M}$ and ${R}$ in the last step.\footnote{We will do this substitution in all final expressions for the self-force. However, in the intermediate calculations, we still have to use ${M=M_0+M_1\epsilon+\dots}$, ${R=R_0+R_1\epsilon+\dots}$, and to distinguish ${M}$, ${R}$ and ${M_0}$, ${R_0}$. Mostly, this should not cause confusion, and it improves the readability of the final results.}
The force is expressed in terms of the Schwarzschild coordinate ${r}$ of the particle with the help of~\eqref{partposr}.

We see that in the leading order, we have obtained just a term that does not depend on the charge of the particle. It should be compared with the gravitational force acting on the particle in the static frame. However, first we have to identify the rest mass of the particle. The mass ${m}$ of the massive object in the limiting procedure has the meaning of the asymptotic mass \cite{Alekseev:2011nj}. For a point particle, the asymptotic mass ${\hat{m}}$ is the energy evaluated at infinity and thus it is related to the rest mass ${\mo}$ as \begin{equation}\label{momrel}
    \hat{m} = \mo \sqrt{1-\frac{2M}{r}}\,.
\end{equation}
Substituting into the expansion of the tension, we find that the first order term of the external force needed to support the particle is
\begin{equation}\label{gravforce}
    F_{\text{ext\ }1} \equiv \tau_1 = \frac{\mo M}{r^2}\Bigl(1-\frac{2M}{r}\Bigr)^{\!-\frac12}\,,
\end{equation}
which is exactly the force acting against the static gravitational force, cf.\ the first term in~\eqref{classforce}.

If we were not sure about the interpretation of the mass parameter ${\hat{m}}$, we could reverse the argument. The leading term should reproduce the gravitational force, and from that we obtain the relation \eqref{momrel} between ${\hat{m}}$ and the rest mass ${\mo}$.

Returning to the expansion \eqref{tensionexp} of the tension, we see that we obtained the self-force contributions only in the second order. The first term in order ${\epsilon^2}$, the term depending on the mass ${\hat{m}}$, is related to the gravitational self-force. The second term proportional to ${q_1^2}$, is related to the electromagnetic self-force.

There is an important difference between these two terms. The electromagnetic self-force does not depend on the details of the limiting procedure hidden in coefficients ${M_1}$ and ${R_1}$. On the contrary, the gravitational self-force does depend on these details. We thus obtain that the external force on the point particle needed to balance the electromagnetic self-force is
\begin{equation}\label{EMselfforce}
    F_{\text{ext EM\ }2} = -\frac{\qo^2 M}{r^3} \frac{1-\frac{M}{r}}{1-\frac{2M}{r}}\,,
\end{equation}
where, for aesthetic reasons,\footnote{%
Here, ${\qo}$ does not refer to a ``rest'' charge similarly to the rest mass~${\mo}$, but it just indicates that it is an intrinsic characteristic of the test particle.}
we changed ${q_1\to\qo}$. This result does not depend on further details of the limit.

Surprisingly, it is not the same as the standard electromagnetic self-force obtained earlier \cite{Vilenkin:1979,Smith:1980tv}, cf.\ the second term in~\eqref{classforce}. It coincides with the classical result for a large radius, ${r\gg M}$, but it differs closer to the horizon. This difference is due to fact that we have consistently incorporated the backreaction of the agent causing the force, namely, of the strut, on the spacetime. The electromagnetic field of the massive object is influenced by the presence of the strut in the fully interacting system. And this influence modifies the resulting force in the limit. The effect is bigger when the strut is short and its energy density and tension are large. This corresponds exactly to the case when the point particle is close to the horizon.

Finally, we should investigate the gravitational self-force. We have already observed that, in contrast to the electromagnetic self-force, it depends on the choice of the family of spacetimes parametrized by the small parameter $\epsilon$; namely, it depends on what is held fixed in the limit $\epsilon\to 0$. The main result could be as follows: the gravitational self-force on the point particle has a well-defined meaning only after an explicit description of how the limit of a point particle is obtained.

In order to demonstrate this type\footnote{This is the type (iii) freedom discussed in the Introduction, see \cite{Pound:2015fma}.} of freedom, we choose several reasonable limiting procedures and show the corresponding self-forces.

\vspace{-2ex}
\subsubsection*{Constant mass ${M}$ and separation between centers}
\vspace{-2ex}

Mathematically, the simplest choice is to assume that the mass ${M(\epsilon)}$ and the separation parameter ${R(\epsilon)}$ do not change during the limit. This means
\begin{equation}\label{MRconst}
    M_1=0\,,\quad R_1=0\,,
\end{equation}
leading to
\begin{equation}\label{sfMRconst}
    F_{\text{ext gr\ }2}
       = \frac{2\hat{m}^2 M^2}{r^4\bigl(1-\frac{2M}{r}\bigr)^2}
       = \frac{2\mo^2 M^2}{r^4\bigl(1-\frac{2M}{r}\bigr)}\,,
\end{equation}
where the last formula is expressed in terms of the rest mass ${\mo}$ using \eqref{momrel}. The corresponding self-force is thus attractive; i.e., it points in the opposite direction to the electromagnetic self-force. It also decreases faster with the radius.

\vspace{-2ex}
\subsubsection*{Constant mass ${M}$ and separation between horizons}
\vspace{-2ex}

A more natural choice may be to keep  the coordinate separation between the horizons of two black holes fixed,
\begin{equation}
\begin{split}
    &R-\Sigma-\sigma \\
    &\quad= (R_0{-}M_0) + (R_1{-}M_1{-}\sigma_1)\,\epsilon +\dots = \text{const}\,,
\end{split}
\end{equation}
with ${\sigma_1}$ given by \eqref{sigma1}. Assuming also a constant mass, ${M=\text{const}}$, we obtain
\begin{equation}\label{MRHhconst}
    M_1=0\,,\quad R_1=\sigma_1\,,
\end{equation}
and for the force
\begin{equation}\label{sfMRHhconst}
    F_{\text{ext gr\ }2} = -\frac{2\mo\sqrt{\mo^2{-}\qo^2} M}{r^3}
      +\frac{2\mo\bigl(\mo{-}\sqrt{\mo^2{-}\qo^2}\bigr)M^2}{r^4\bigl(1-\frac{2M}{r}\bigr)}\,.
\end{equation}
We see that the gravitational self-force is influenced by the charge of the particle in this case. It is well defined only for ${\mo^2>\qo^2}$, which is related to the fact that we have assumed the existence of both horizons, i.e., that the massive object in the limiting process is a black hole. The first term is dominant for large ${r}$, and it also remains for an uncharged particle, ${\qo=0}$, when
\begin{equation}\label{sfMRHhconstq0}
    F_{\text{ext gr\ }2} = -\frac{2\mo^2 M}{r^3}\,.
\end{equation}
The self-force is repulsive from the black hole in this case.

\vspace{-2ex}
\subsubsection*{Constant total mass and separation between centers}
\vspace{-2ex}

Requiring the total mass ${\mathcal{M}=M+m}$ and ${R}$ constant, we get
\begin{equation}\label{totMRconst}
   M_1 = -\hat{m}\,,\quad R_1=0\,,
\end{equation}
and for the force we obtain a rather simple expression
\begin{equation}\label{sftotMRconst}
   F_{\text{ext gr\ }2} = -\frac{2\mo^2}{r^2}\,.
\end{equation}
Surprisingly, it does not depend on the mass ${M}$ of the big black hole; it depends on ${r}$ by the inverse square law. Thus, it decreases at the same rate as the standard gravitational force~\eqref{gravforce}.

\vspace{-2ex}
\subsubsection*{Constant entropy ${S}$ and thermodynamic length}
\vspace{-2ex}

Assuming that the massive object is a black hole, we can require the entropy of the big black hole ${S}$ and the thermodynamic length ${\ell_{\srm{Hh}}}$ to be constant during the limit. Expanding the first expression in \eqref{TDQ0} and \eqref{ellHhQ0}, we obtain
\begin{align}
  S &= 4\pi M_0 + 8\pi M_0\Bigl(M_1+\frac{\hat{m} M_0}{R_0{+}M_0}\Bigr)\,\epsilon + \dots\,,
  \label{Sexp}\\
  \ell_{\srm{Hh}} &= (R_0{-}M_0)
    +\Bigl(R_1{-}M_1{-}\sigma_1{-}\frac{4\hat{m}M_0}{R_0{+}M_0}\Bigr)\,\epsilon+\dots\,.
  \label{ellHhexp}
\end{align}
Requiring the first order terms to vanish, we get
\begin{equation}\label{SellHhconst}
    M_1=-\frac{\hat{m} M_0}{R_0{+}M_0}\,,\quad
    R_1=\sigma_1+\frac{3\hat{m}M_0}{R_0{+}M_0}\,.
\end{equation}
Substituting into the formula \eqref{tensionexp}, we get an unimpressive result
\begin{equation}\label{sfSellHhconst}
\begin{split}
    F_{\text{ext gr\ }2} = -&\frac{6\mo^2 M}{r^3}\frac{1-\frac{4M}{3r^2}}{1-\frac{2M}{r}}\\
      +&\frac{2\mo\bigl(\mo{-}\sqrt{\mo^2{-}\qo^2}\bigr)M}{r^3}\frac{1-\frac{M}{r}}{1-\frac{2M}{r}}\,.
\end{split}
\end{equation}

Assuming that the massive object is a naked singularity, we require that the thermodynamic length ${\ell_{\srm{Hs}}}$ given by \eqref{ellHsQ0} is constant. Its expansion is
\begin{equation}\label{ellHsexp}
  \ell_{\srm{Hs}} = (R_0{-}M_0)
    +\Bigl(R_1{-}M_1{+}\hat{m}{-}\frac{4\hat{m}M_0}{R_0{+}M_0}\Bigr)\,\epsilon+\dots\,,
\end{equation}
which yields
\begin{equation}\label{SellHsconst}
    M_1=-\frac{\hat{m} M_0}{R_0{+}M_0}\,,\quad
    R_1=\hat{m}\frac{2M_0{-}R_0}{R_0{+}M_0}\,.
\end{equation}
For the force, we obtain
\begin{equation}\label{sfSellHsconst}
   F_{\text{ext gr\ }2} = \frac{\mo^2 M}{r^3}\Bigl(1-\frac{2M}{r}\Bigr)\,.
\end{equation}
It is worth noting that the gravitational self-force is again attractive in this case. It is also independent of the charge of the particle.

\vspace{-2ex}
\subsubsection*{Constant temperature ${T}$ and thermodynamic length}
\vspace{-2ex}

Similarly to the entropy, we can keep constant the temperature (the surface gravity) of the big black hole. Its expansion reads
\begin{equation}\label{Texp}
    T = \frac{1}{8\pi M_0} - \frac{1}{8\pi M_0}\biggl(
      \frac{2\hat{m}}{R_0{+}M_0}+\frac{M_1}{M_0}\biggr)\,\epsilon
      +\dots\,,
\end{equation}

For the limit of a small black hole, we require the thermodynamic length ${\ell_{\srm{Hh}}}$ to be constant. This means that the first order terms in expansions \eqref{Texp} and \eqref{ellHhexp} must vanish, which yields
\begin{equation}\label{TellHhconst}
    M_1=-\frac{2\hat{m} M_0}{R_0{+}M_0}\,,\quad
    R_1=\sigma_1+\frac{2\hat{m}M_0}{R_0{+}M_0}\,.
\end{equation}
The force turns out to be
\begin{equation}\label{sfTellHhconst}
\begin{split}
    F_{\text{ext gr\ }2} = -&\frac{4\mo^2 M}{r^3}\frac{1-\frac{M}{r}}{1-\frac{2M}{r}}\\
      +&\frac{2\mo \bigl(\mo{-}\sqrt{\mo^2{-}\qo^2}\bigr)M}{r^3}\frac{1-\frac{M}{r}}{1-\frac{2M}{r}}\,.
\end{split}
\end{equation}

In the case of a naked singularity limit, we require the thermodynamic length ${\ell_{\srm{Hs}}}$ to be fixed. From \eqref{Texp} and \eqref{ellHsexp} it follows that
\begin{equation}\label{TellHsconst}
    M_1=-\frac{2\hat{m} M_0}{R_0{+}M_0}\,,\quad
    R_1=-\hat{m}\frac{R_0{-}M_0}{R_0{+}M_0}\,.
\end{equation}
Surprisingly, all contributions to the gravitational self-force cancel each other in this case,
\begin{equation}\label{sfTellHsconst}
   F_{\text{ext gr\ }2} = 0\,.
\end{equation}

\vspace{-2ex}
\subsubsection*{Constant mass ${M}$ and proper length between horizons}
\vspace{-2ex}

As the last example, we discuss the limit of a small black hole with mass ${M}$ and the proper length between black-hole horizons ${L_{\srm{Hh}}}$ fixed. The expansion of the proper length \eqref{LHhapp} is discussed in Appendix~\ref{apx:PropLength},
\begin{equation}\label{LHhexp}
\begin{split}
    &L_{\srm{Hh}} = \sqrt{R_0^2{-}M_0^2}+2M_0\arctanh\sqrt{\frac{R_0{-}M_0}{R_0{+}M_0}}\\
      &\;-\Biggl(\hat{m}\frac{4 M_0}{\sqrt{R_0^2{-}M_0^2}}
      +(M_1{-}R_1)\frac{R_0{+}M_0}{R_0{-}M_0}\\
      &\quad+2\biggl(\hat{m}\frac{R_0^2{+}3M_0^2}{R_0^2{-}M_0^2}-M_1\biggr)
         \arctanh\sqrt{\frac{R_0{-}M_0}{R_0{+}M_0}}\\
      &\quad+\hat{m}\frac{R_0^2{+}3M_0^2}{R_0^2{-}M_0^2}
         \log\frac{\sigma_1 M_0\, \epsilon}{4(R_0^2{-}M_0^2)}
      \Biggr)\,\epsilon+\dots\,.
\end{split}
\end{equation}
A new feature here is that the expansion contains logarithmic terms ${\log\epsilon}$. This reflects the nonanalytic dependence of the proper length on the expansion parameter. However, one can still require that the linear terms of expansion of ${M}$ and ${L_{\srm{Hh}}}$ vanish, yielding
\begin{equation}\label{MLHhconst}
\begin{gathered}
    M_1=0\,,\\
\begin{aligned}
    R_1&=\frac{4\hat{m}M_0}{R_0{+}M_0}
       +\hat{m}\log\frac{\sigma_1 M_0\, \epsilon}{4(R_0^2{-}M_0^2)}\\
    &+\frac{2\hat{m}\bigl(R_0^2{+}3M_0^2)}{(R_0{+}M_0)\sqrt{R_0^2{-}M_0^2}}
       \arctanh\sqrt{\frac{R_0{-}M_0}{R_0{+}M_0}}
    \,.
\end{aligned}
\end{gathered}
\end{equation}
Substituting to the tension \eqref{tensionexp} gives a complicated expression for the force
\begin{equation}\label{sfMLHhconst}
\begin{split}
   &F_{\text{ext gr\ }2} = -\frac{2\mo^2 M}{r^4\bigl(1{-}\frac{2M}{r}\bigr)}\Biggl(
      M+2M\Bigl(1{-}\frac{2M}{r}\Bigr)\\
      &\quad+2\frac{r-M}{\sqrt{1{-}\frac{2M}{r}}}\biggl(1{-}\frac{2M}{r}{+}\frac{4M^2}{r^2}\biggr)
         \arctanh\sqrt{1{-}\frac{2M}{r}}\\
      &\quad+(r-M)\log\frac{\sqrt{\mo^2{-}\qo^2} M\,\epsilon}{4r^2\sqrt{1{-}\frac{2M}{r}}}
   \Biggr)\,.
\end{split}\raisetag{9ex}
\end{equation}

We derived this expression mainly because it shows that the physically well-motivated condition of the fixed proper distance can lead to logarithmic divergences in the self-force. Of course, the self-force is of the second order in ${\epsilon}$, so the logarithmic term is of the type ${\epsilon^2\log\epsilon}$ which is not a real divergence. But it still documents a broad range of behavior of the self-force, depending on the limiting procedure.

A similar analysis can be made in the naked singularity case, using the proper length ${L_{\srm{Hs}}}$ given by \eqref{LHs}. The expansion of the elliptic integrals is even more problematic, and the result is not a simple expression. It contains logarithmic terms, and it depends on the charge of the particle. Because it does not offer anything qualitatively new, we skip it here.


\subsection{Limit ${m,\,q\to 0}$ with ${m\ll q}$}
\label{ssc:limitmlessq}

By discussing various limiting procedures, we have clearly demonstrated that the gravitational self-force in this approximation is not uniquely defined. However, it raises the question of the well-definiteness of the electromagnetic force, which is of the same order. Can one take the expression \eqref{EMselfforce} seriously if it should be combined with a non-unique expression for the gravitational contribution? One could argue that the electromagnetic self-force is identified by its dependence on the square ${\qo^2}$ of the test charge. However, we have seen that the gravitational self-force can also depend on the charge.

However, we can modify our approximation by assuming that the mass ${m}$ of the massive object is much smaller than its charge ${q}$. This implies that the massive object must be modeled by a naked singularity. Although this can raise suspicions, the values of the charge and mass of elementary particles satisfy the condition ${\mo<|\qo|}$. We implement this  by changing the expansion \eqref{mqExpansion} as~follows:
\begin{equation}\label{mg<qExpansion}
\begin{aligned}
    m(\epsilon) &= \hat{m}\,\epsilon^2\,,\\
    q(\epsilon) &= q_1\,\epsilon + q_2\, \epsilon^2 + \dots\,,\\
    M(\epsilon) &= M_0 + M_1\, \epsilon + \dots\,,\\
    R(\epsilon) &= R_0 + R_1\, \epsilon + \dots\,.\\
\end{aligned}
\end{equation}
The mass ${m}$ thus approaches zero faster than the charge~${q}$.

Not surprisingly, the expansion of the tension \eqref{tensionexp} changes to
\begin{equation}\label{tensionexpm<q}
    \tau = \frac{\hat{m}M}{r^2}\Bigl(1-\frac{2M}{r}\Bigr)^{\!-1}\epsilon^2
     -\frac{q_1^2 M}{r^3} \frac{1-\frac{M}{r}}{1-\frac{2M}{r}}\,\epsilon^2
     + \dots \,.
\end{equation}
It defines the force needed to support the test point particle at a static position as
\begin{equation}\label{selfforcem<q}
    F_{\text{ext}} =
     \frac{\mo M}{r^2}\Bigl(1-\frac{2M}{r}\Bigr)^{\!-\frac12}
     -\frac{\qo^2 M}{r^3} \frac{1-\frac{M}{r}}{1-\frac{2M}{r}}\,,
\end{equation}
where we again introduced the rest mass ${\mo}$ by \eqref{momrel} and the symmetric notation for the charge, ${\qo\equiv q_1}$.

Clearly, the first term compensates the gravitational force in the static frame, and the second term is the electromagnetic self-force derived above in \eqref{EMselfforce}. Further corrections corresponding to the gravitational self-force are now of higher order, and we ignore them. In this context it makes sense to discuss the electromagnetic self-force alone. The result \eqref{selfforcem<q} should thus be compared with the classical result \eqref{classforce}. As discussed above, we have obtained a modification of the self-force near the horizon due to the gravitational influence of the strut on the electromagnetic field.


\vspace{-1ex}

\section{Summary}\label{sc:Summary}

\vspace{-1ex}

To obtain a better understanding of the nature of the point-like particle approximation, we investigate a fully interacting system of a big neutral black hole with an extended charged massive object nearby. The massive object is modeled by a small black hole or a naked singularity, which corresponds at the limit to the particle with mass and charge satisfying ${\mo\gtrsim\qo}$ or ${\mo\lesssim\qo}$, respectively. This system obeys the full Einstein--Maxwell equations; the massive object is kept in equilibrium above the black hole by a strut with a linear tension which balances the gravitational and electromagnetic interaction with the black hole.

By shrinking the massive object to a point, we obtain the Schwarzschild spacetime with a test point-like charged particle supported on the static orbit by a test strut. The tension of the strut defines the external force needed to balance the gravitational force of the black hole and the gravitational and electromagnetic~self-forces.

When we choose the limiting procedure such that the mass and the charge of the massive object approach zero in the same order, we find that the leading term of the tension of the strut corresponds to the standard gravitational force~\eqref{gravforce} of the black hole acting on the particle. In the next order we find that the tension also compensates the electromagnetic and gravitational self-forces. The electromagnetic self-force is given by expression~\eqref{EMselfforce}. It is independent of any further details of the limiting procedure.

The gravitational self-force, on the other hand, depends on the details of what is kept fixed while taking the limit of small mass and charge of the test particle. We have demonstrated that, by a suitable choice of the limit, one can achieve very different results for the self-force: it can be attractive or repulsive, cf.\ \eqref{sfMRconst} vs.\ \eqref{sfMRHhconst}; it may or may not depend on the charge, cf.~\eqref{sfSellHhconst} vs.\ \eqref{sfSellHsconst}; and it can be independent of the mass of the black hole, see \eqref{sftotMRconst}. It may even completely vanish, cf.~\eqref{sfTellHsconst}, or it can contain terms logarithmic in the expansion parameter, see~\eqref{sfMLHhconst}. It is clear that one has to choose very well-founded physical reasons for how to perform the limiting procedure in order to obtain a trustworthy and unambiguous result.

If we choose the particle mass to approach zero in higher order than the charge, i.e.\ ${\mo\ll\qo}$, we obtain to leading order the standard gravitational force and the electromagnetic self-force, together given by formula \eqref{selfforcem<q}. The gravitational self-force is of higher order now and can be ignored.

The electromagnetic self-force \eqref{selfforcem<q} obtained in our model differs from the classical result \eqref{classforce} in a domain near the horizon. The reason for this difference is that we have taken into account the influence of the strut (the agent supporting the massive object) on the surrounding geometry and thus also on the electromagnetic field. The effect is strong for a short strut with large linear energy density and tension, i.e., exactly when the massive object is near the horizon. As a consequence, our formula for the electromagnetic self-force diverges on the horizon.

When considering the result \eqref{selfforcem<q}, one can easily check that there exists a self-floating solution when the electromagnetic self-force compensates the gravitational force and the strut is not needed (it has vanishing energy and tension). However, as for a similar situation discussed for the classical electromagnetic self-force \cite{Smith:1980tv}, parameters of such a solution are unphysical. This happens for the mass of the black hole and the position of the particle being of the order of the ``classical radius'' ${\frac{\qo^2}{\mo}}$ of the point particle. In this regime quantum effects spoil the validity of the classical theory which we are assuming.


\vspace{-1ex}

\acknowledgements


We are grateful to George Alekseev for clarifying some nontrivial properties and the structure of singularities of the double charged black-hole solution \cite{Alekseev:2007re,Manko:2007hi}.

P.K. was supported by Czech Science Foundation Grant No.~19-01850S. The work was done under the auspices of the Albert Einstein Center for Gravitation and Astrophysics, Czech Republic. P.K. also thanks the University of Alberta for hospitality.
A.Z. thanks the Natural Sciences and Engineering Research Council of Canada and the Killam Trust for the financial support and appreciates the hospitality and support of the Institute of Theoretical Physics of the Charles University in Prague.



\appendix


\section{Metric functions}
\label{apx:mtrcfcs}


The metric functions ${f}$ and ${h}$ and the potential $\Phi$ have been specified in \eqref{fh} and \eqref{Phi} using auxiliary functions $\mathcal{A}$, $\mathcal{B}$ and $\mathcal{C}$. These functions read \cite{Alekseev:2007re,Alekseev:2011nj,Manko:2007hi,Manko:2008gb}
\be\begin{split}\label{Adef}
  \mathcal{A}={}&\Sigma\sigma\bigl[
    \nu(R_{\sss+}{+}R_{\sss-})(r_{\sss+}{+}r_{\sss-})
    +4\kap(R_{\sss+} R_{\sss-}{+}r_{\sss+} r_{\sss-})\bigr]\\
    &-(\mu^2\nu{-}2\kap^2)(R_{\sss+}{-}R_{\sss-})(r_{\sss+}{-}r_{\sss-})\,,
\end{split}\ee
\be\begin{split}\label{Bdef}
  \mathcal{B}=2&\Sigma\sigma\bigl[
    (\nu m{+}2\kap M)(R_{\sss+}{+}R_{\sss-})
    +(\nu M{+}2\kap m)(r_{\sss+}{+}r_{\sss-})\bigr]\\
    -2&\sigma\bigl[\nu\mu(Q{-}\mu)-2\kap(R M{-}\mu q{-}\mu^2)\bigr](R_{\sss+}{-}R_{\sss-})\\
    -2&\Sigma\bigl[\nu\mu(q{+}\mu)-2\kap(R m{+}\mu Q{-}\mu^2)\bigr](r_{\sss+}{-}r_{\sss-}),
\end{split}\raisetag{6.5ex}\ee
\be\begin{split}\label{Cdef}
  \mathcal{C}=2&\Sigma\sigma\bigl[
    \bigl(\nu(q{+}\mu)+2\kap(Q{-}\mu)\bigr)(R_{\sss+}{+}R_{\sss-})\\
    &\quad+\bigl(\nu(Q{-}\mu)+2\kap(q{+}\mu)\bigr)(r_{\sss+}{+}r_{\sss-})\bigr]\\
    -2&\sigma\bigl[\nu\mu M +2\kap(\mu m{-}RQ{+}\mu R)\bigr](R_{\sss+}{-}R_{\sss-}) \\
    -2&\Sigma\bigl[\nu\mu m +2\kap(\mu M{+}Rq{+}\mu R)\bigr](r_{\sss+}{-}r_{\sss-})\,.
\end{split}\ee
The quantities involved here have been introduced in Sec.~\ref{ssc:geometry}.


\section{Expansion of various quantities}
\label{apx:expansions}

A derivation of the tension expansion in the limit \eqref{mqExpansion} is straightforward, but tedious. First, we list some intermediate expansions for the involved quantities.

We start with the expansion of the constant ${\mu}$ defined in \eqref{constsdef},
\begin{widetext}
\begin{equation}\label{muexp}
   \mu = \frac{M_0 q_1}{R_0+M_0}\,\epsilon
    +\biggl( \frac{M_1 q_1+M_0 q_2}{R_0+M_0}
    -\frac{M_0 q_1\, (\hat{m}+M_1+R_1)}{({R_0+M_0})^2}
    \biggr)\,\epsilon^2 + \dots\,.
\end{equation}
Next, the half-length ${\sigma}$, cf.~\eqref{sigmas}, is
\begin{equation}\label{sigmaexp}
    \sigma = \sigma_1\, \epsilon - \frac{q_1}{\sigma_1}\,
    \frac{q_1\,(\hat{m}M_0 + M_0 R_1 - M_1 R_0)+q_2\,(R_0^2-M_0^2)}{(R_0+M_0)^2}\,\epsilon^2
    +\dots\,,
\end{equation}
\end{widetext}
where
\begin{equation}\label{sigma1def}
    \sigma_1 = \sqrt{\hat{m}^2-q_1^2\, \frac{R_0-M_0}{R_0+M_0}}
       = \sqrt{\mo^2-\qo^2}\,\sqrt{1-\frac{2M}{r}}\;.
\end{equation}
The last formula is just expressed in terms of the rest mass ${\mo}$, charge ${\qo\equiv q_1}$ and the Schwarzschild coordinate ${r=R_0+M_0}$.

Finally, for the constants ${\nu}$ and ${\kap}$, cf.~\eqref{constsdef}, the expansions are
\begin{gather}
  \nu = \bigl(R_0^2-M_0^2\bigr) + 2 \bigl(R_0 R_1 - M_0 M_1\bigr)\,\epsilon + \dots\,,
    \label{nuexp}\\
  \kap = \hat{m} M_0\,\epsilon
    + \biggl( \hat{m} M_1 - \frac{q_1^2\,M_0 R_0 }{(R_0+M_0)^2}\biggr)\,\epsilon^2+ \dots\,.
    \label{kapexp}
\end{gather}


\section{Proper length between the black hole and a massive object}
\label{apx:PropLength}

The proper length of the strut along the symmetry axis is
\be\label{Lcalc}
    L =\int_{\text{strut}}\bigl(h\,f^{-1/2}\bigr)\big|_{\rho=0}\,dz\,.
\ee
The metric function ${h}$ is constant on the axis and on the strut it takes the value ${h_{\mathrm{o}}}$ given by
\eqref{honaxis}. The metric function ${f}$ on the strut takes the form
\begin{equation}\label{fonaxis}
    f = \frac{\bigl((z-z_H)^2-\Sigma^2\bigr)\bigl((z-z_h)^2-\sigma^2\bigr)}
          {((z-z_H+M)(z_h-z+m) - Q q )^2}
\end{equation}

For the case of two black holes, the integration in \eqref{Lcalc} runs between the horizons ${z\in(z_H+\Sigma,z_h-\sigma)}$ and we get
\be\label{LHhint}
   L_{\srm{Hh}} =h_{\mathrm{o}} \int_{z_H+\Sigma}^{z_h-\sigma}  \frac{(z-z_H+M)(z_h-z+m) - Q q }
    {\sqrt{\bigl((z-z_H)^2-\Sigma^2\bigr)\bigl((z-z_h)^2-\sigma^2\bigr)}}\,dz\,.
\ee
After some substitutions and manipulations, and using integral tables, one can derive the result in terms of elliptic integrals,
\be\begin{split}\label{LHhapp}
  L_{\srm{Hh}}=\frac{h_{\mathrm{o}}}{\xi}&\bigg(\xi^2\,\mathsf{E}(k)+
     4m\Sigma\,\mathsf{\Pi}(\alpha^2,k)+4M\sigma\,\mathsf{\Pi}(\Alpha^{\!2},k)\\
  &+2\bigl(M m{-}Q q{-}M\sigma{-}m\Sigma{-}\Sigma\sigma\bigr)\,\mathsf{K}(k)
     \bigg)\,.
\end{split}\raisetag{7ex}
\ee
where
\begin{equation}\label{elintpar}
\begin{aligned}
  \xi^2&=R^2-(\Sigma-\sigma)^2\,, &
  \alpha^2&=\frac{R-\Sigma-\sigma}{R+\Sigma-\sigma}\,,
  \\
  k^2&=\frac{R^2-(\Sigma+\sigma)^2}{R^2-(\Sigma-\sigma)^2}\,,&
  \Alpha^{\!2}&=\frac{R-\Sigma-\sigma}{R-\Sigma+\sigma}\,.
\end{aligned}
\end{equation}

In the test charge limit $\sigma\to 0$ and, hence,
\be
k^2\to 1, \hskip 1.5cm \Alpha^{\!2}\to 1.
\ee
The expansion at $k=1$ of the functions $\mathsf{E}(k)$ and $\mathsf{K}(k)$ does not pose any problems. But the expansion of the elliptic integrals $\mathsf{\Pi}(\alpha^{\!2},k)$ and especially $\mathsf{\Pi}(\Alpha^{\!2},k)$  in this limit is less evident.

First of all we rewrite \eqref{LHhapp} using the following  property of the elliptic integrals (see Eq.\ (19.7.9) at \cite{Olver:2010})
\be
  \sigma \mathsf{\Pi}(\Alpha^{\!2},k)+\Sigma \mathsf{\Pi}(\alpha^2,k)
  =\frac{1}{2}(R+\Sigma+\sigma)\,\mathsf{K}(k)
\ee
This makes it possible to rewrite \eqref{LHhapp} in an equivalent non-symmetrical form,
\be\begin{split}\label{LHhasym}
   L_\srm{Hh} &= \frac{h_{\mathrm{o}}}{\xi}
      \biggl( \xi^2\,\mathsf{E}(k)-4\Sigma(M-m)\mathsf{\Pi}(\alpha^2,k)\\
      &+2\bigl(MR{+}Mm{-}Qq{-}\Sigma\sigma{+}\Sigma(M{-}m)\bigr)\mathsf{K}(k)\biggr)\,.
\end{split}\ee
This form is much better suited for the series expansion at small $m$ and $q$.
Then we use the following  representation
\be
\begin{split}\label{Pialpha}
    &\mathsf{\Pi}(\alpha^2,k)=\mathsf{K}(k)\\
    &-\frac{\alpha}{\sqrt{1{-}\alpha^2}\sqrt{k^2{-}\alpha^2}}\Bigl(\mathsf{E}(k) \mathsf{F}(\beta,k)-\mathsf{K}(k) \mathsf{E}(\beta,k)\Bigr)\,,
    \end{split}
\ee\\[0.5ex]
where ${\sin \beta=\frac{\alpha}{k}}$. This identity is valid for all ${0<k<1}$ and ${0<\alpha<k}$, and it is convenient to find the series expansion at $k=1$ since the expansions of incomplete elliptic integrals $\mathsf{E}(\beta,k)$ and $\mathsf{F}(\beta,k)$ at $k=1$ are well known.

To write down these expansions, it is useful to introduce the quantity $k'=\sqrt{1-k^2}$ and compute the series at $k'=0$. The list of necessary expansions is
\begin{widetext}\noindent
\be
\mathsf{K}(k)=-\ln\frac{k'}{4}-\frac{1}{4}\Bigl(\ln\frac{k'}{4}+1\Bigr)\,k'^2
-\frac{9}{64}\Bigl(\ln\frac{k'}{4}+\frac{7}{6}\Bigr)\,k'^4
+\mathcal{O}\bigl(k'^6\bigr)\,,
\ee
\be
\mathsf{E}(k)=1-\frac{1}{2}\Bigl(\ln\frac{k'}{4}+\frac{1}{2}\Bigr)\,k'^2
-\frac{3}{16}\Bigl(\ln\frac{k'}{4}+\frac{13}{12}\Bigr)\,k'^4
+\mathcal{O}\bigl(k'^6\bigr)\,,
\ee
\be
\mathsf{F}(\beta,k)=\arctanh\alpha+\frac{1}{4}\Bigl(\arctanh\alpha+\frac{\alpha}{1-\alpha^2}\Bigr)\,k'^2
+\frac{9}{64}\biggl(\arctanh\alpha+\frac{\alpha(5-3\alpha^2)}{3(1-\alpha^2)^2}\biggr)\,k'^4+\mathcal{O}\bigl(k'^6\bigr)\,,
\ee
\be
\mathsf{E}(\beta,k)=\alpha+\frac{1}{2}\arctanh\alpha\;k'^2
+\frac{3}{16}\Bigl(\arctanh\alpha+\frac{\alpha}{1-\alpha^2}\Bigr)\,k'^4
+\mathcal{O}\bigl(k'^6\bigr)\,.
\ee
Using the identity \eqref{Pialpha} we get
\be
\begin{split}
\mathsf{\Pi}(\alpha^2,k)=-&\frac{\ln\frac{k'}{4}+\alpha\arctanh\alpha}{1-\alpha^2}
-\frac{1+(1+\alpha^2)\ln\frac{k'}{4}+2\alpha\arctanh\alpha}{4(1-\alpha^2)^2}\,k'^2\\
-&\frac{1}{128(1-\alpha^2)^3}\Bigl(6(3+6\alpha^2-\alpha^4)\ln\frac{k'}{4}
+\bigl(48\alpha\arctanh\alpha+21+12\alpha^2-5\alpha^4\bigr)
\Bigr)\,k'^4+\mathcal{O}\bigl(k'^6\bigr)\,.
\end{split}
\raisetag{14ex}
\ee
\end{widetext}
Using the expansion \eqref{mqExpansion} of the spacetime parameters in formulas \eqref{honaxis} \eqref{elintpar}, substituting these in the series expansions above, and putting them all together in \eqref{LHhasym}, we eventually obtain the result \eqref{LHhexp}.

The case with a naked singularity is a bit more involved, and here we present only the formula for the case we are interested in, namely, for a naked singularity of mass ${m}$ and charge ${q}$ near a neutral black hole of mass ${M}$. In this case one has to integrate over ${z}$ from the horizon up to the naked singularity ${z\in(z_H+M,z_h+m)}$, which gives the expression
\be\begin{split}\label{LHs}
  L_{\srm{Hs}}&=\frac{h_{\mathrm{o}}}{\xi}\bigg(
     \xi^2\,\mathsf{E}(\psi,k)-\xi\sqrt{m^2+\tilde\sigma^2}\sqrt{\frac{R-M+m}{R+M+m}}\\[1ex]
  &-4M(M{-}m)\mathsf{\Pi}(\psi,\alpha^2,k)+2M(R{+}M{-}i\tilde\sigma)\mathsf{F}(\psi,k)\bigg).
\end{split}\raisetag{8.5ex}
\ee
Here ${\xi}$, ${k^2}$, ${\alpha^2}$, and ${h_{\mathrm{o}}}$ are again given by \eqref{elintpar} and \eqref{honaxis}, but with ${\Sigma=M}$ and an imaginary ${\sigma=i\tilde\sigma}$, where a real $\tilde\sigma$ is
\begin{equation}\label{sigmabar}
    \tilde\sigma = \sqrt{q^2+2\mu q-m^2}\,.
\end{equation}
The constant ${\psi}$ is given by
\be\label{psidef}
\sin\psi=\sqrt{\frac{(R+M-i\tilde\sigma)(R-M+m)}{(R-M-i\tilde\sigma)(R+M+m)}}\,.
\ee
The expression \eqref{LHs} contains complex arguments, nevertheless, one can show that $L$ is real, as it should be.




%

\end{document}